\documentclass[aps,pra,twocolumn,longbibliography,showpacs]{revtex4-1}
\usepackage{amssymb, graphicx}
\usepackage{enumerate}
\usepackage{amsmath,bm}
\usepackage{dcolumn}
\usepackage{mathrsfs}
\usepackage[latin1]{inputenc}
\newcommand{\melement}[3]{\ensuremath{\left\langle #1 \left|#2\right|#3\right\rangle}}%
\newcommand{\degree}{\ensuremath{^\circ}}%
\newcommand{\ie}{i.\,e.}%
\newcommand{\tensornote}[1]{\ensuremath{\underline{\underline{#1}}}}
\newcommand{\operator}[1]{\ensuremath{\widehat{#1}}}

\makeatletter
\newcommand{\ket}[1]{\left|#1\right\rangle}
\newcommand{\bra}[1]{\left\langle #1\right|}

\newcommand{\expected}[1]{\left\langle #1\right\rangle}
\newcommand{\power}[1]{\ensuremath{^{#1}}}
\newcommand{\aarphys}{\affiliation{Department of Physics and Astronomy, Aarhus University, 8000  Aarhus C, Denmark}}%

\begin{document}

\title{Torsional and rotational coupling in non-rigid molecules}

\author{Juan J.\ Omiste}\aarphys%\granada

\author{Lars Bojer Madsen}\aarphys

\date{\today}
\begin{abstract} 
  We analyze theoretically the interplay between the torsional and the rotational motion of an aligned biphenyl-like molecule. To do so, we consider a transition between two electronic states with different internal torsional potentials, induced by means of a resonant laser pulse. The change in the internal torsional potential provokes the motion of the torsional wavepacket in the excited electronic state, modifying the structure of the molecule, and hence, its inertia tensor. We find that this process has a strong impact on the rotational wave function, displaying different
 behavior  depending on the electronic states involved and their associated torsional potentials.  We describe the dynamics of the system by 
considering the degree of alignment and  the expectations values of the angular momentum operators for the overall rotation of the molecule. 

\end{abstract}
%37.10.Vz: Mechanical effects of light on atoms, molecules, and ions
%33.15.Hp: Barrier heights (internal rotation, inversion, rotational isomerism, conformational dynamics)
\maketitle
\section{Introduction}
\label{sec:introduction}

The development of new laser sources and experimental techniques has made possible the control and deeper understanding of the electron motion in atoms and nuclei in molecules. The determination of the molecular structure has been a challenge for decades, and now it is accessible to experimental investigations on the atto-, femto- and picosecond time scales by X-ray~\cite{Boll2013,Kupper2014} and electron diffraction~\cite{Peters2011,PhysRevLett.102.213001}, high harmonic generation~\cite{Cireasa2015,Kraus2015}, photo-ionization~\cite{Holmegaard:natphys6,Calegari2014} and Coulomb explosion~\cite{Filsinger2009,Nielsen2012,Christensen2014,Slater2015}. Furthermore, the addition of external electric and magnetic fields allows the control of stereo chemical properties of linear and polyatomic molecules. In particular, the control of translational dynamics~\cite{tokunaga:new_j_phys_11}, orientation and alignment in the laboratory~\cite{Filsinger2009,Nielsen2012} and state~\cite{Kupper2009} or conformers~\cite{Rosch2014} selection have been realized. Recently, many efforts have focused on the transformation between isomers using a combination of non-resonant laser pulses, in particular for molecules formed by two planar groups linked by a bond~\cite{Madsen2009a,Madsen2009,Hansen2012,Christensen2014}. In such molecules a long nanosecond pulse is used to align the main axis of the molecular polarizability tensor either along a given direction or in a plane of the laboratory system. Then, a short picosecond laser pulse or train of pulses interact efficiently with the fixed molecule, kicking the planar groups to or against the coplanar configuration. Theoretically, the description of this process has been tackled using a 2D model, which considers that the molecule is perfectly aligned by the long laser pulse, and the kick pulse only affects the internal rotation and torsion\cite{Christensen2014,Madsen2009a,Madsen2009,Hansen2012,Coudert2015}. The 1D model is a more restrictive approximation which considers only the torsion degrees of freedom~\cite{Ashwell2013a,Coudert2015}. In a strong field, it is in principle possible to design an appropriate laser pulse capable of transferring the whole population to a desired enantiomer~(optical isomer)~\cite{Parker2012}. 

The theoretical studies of the torsional and overall rotation have been carried out for biphenyl-like molecules, which together with ethylene, has been used as  prototypes to highlight the motion in non-rigid molecules~\cite{Merer1973}. We use the term biphenyl-like to describe a molecule formed by two identical rings linked by a bond, whose paradigmatic case is biphenyl. The rings can rotate with respect to each other, giving rise to the torsion. It has been shown theoretically that the interaction with an intense laser pulse has strong implications on the alignment and the torsional dynamics between the two groups~\cite{Coudert2011,Ortigoso2013}. If the laser is strong enough, it is possible to change the internal configuration of the molecule, e.g., to change the chirality in the case of axially chiral molecules. However, this effect is purely theoretical, since the intensities required would lead to ionization or breaking of the molecule before such a transition would occur. In practice, the torsional barrier is much higher than the energy associated with the laser interaction and its impact therefore leads to relatively small fluctuations in the torsional angle~\cite{Coudert2015}.

In this work, we investigate theoretically the connection between the torsional and rotational motion in a biphenyl-like molecule by enabling the torsional relaxation of the rings. By means of a resonant laser we induce a transition between two electronic states~\cite{Baca1979}. As a consequence, the phenyl rings feel a different potential in the final electronic state, inducing their internal torsion. This change in the dynamics leads to a modification of the rotational parameters of the molecule, which depend on the relative position of the rings, and subsequently, initiate the rotational dynamics that may be reflected in an improvement of the alignment of the molecule in the laboratory. The rotational dynamics is understood in terms of the total angular momentum of the entire molecule $\expected{J\power{2}}$ and its projection along the $z$~axis of the molecular fixed frame (MFF) $\expected{K\power{2}}$. The change of $K$ in time means that the molecule undergoes changes in the rotation around the $z$ axis of the MFF which leads to variations on its orientation in space, similar to the classical precession or nutation.

The paper is organized as follows. In Sec.~\ref{sec:the_system}, we describe the physical process under investigation together with the Hamiltonian and the symmetries of the system. Next, in Sec.~\ref{sec:numerical_methods} we explain in detail the numerical methods used and in Sec.~\ref{sec:results} we present the main results. In Sec.~\ref{sec:conclusions_and_outlook} we summarize the conclusions of this work and point to new possibilities. Finally, the Appendices contain details about the polarizability tensor and the calculation of the matrix elements of the Hamiltonian interaction term, as well as the underlying basis functions.

\section{The system}
\label{sec:the_system}

\subsection{Physics of the process}
\label{sec:physics_of_the_process}
\begin{figure}[hbt]
  \begin{center}
    \includegraphics[width=.7\linewidth]{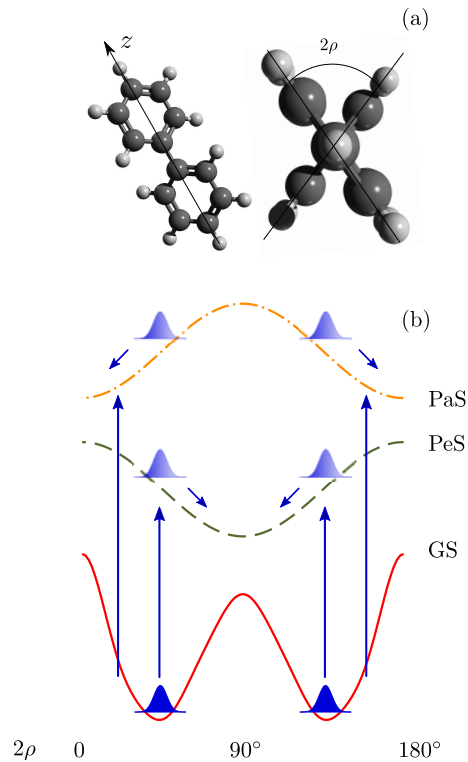}
  \end{center}
  \caption{\label{fig:fig1} (a) Sketch of the biphenyl-like molecule together with the $z$ axis of the MFF and the torsional angle $\rho$ for the molecule seen from the end. (b) Transitions from the ground state potential energy surface (GS) to planar and perpendicular energy surfaces (PaS and PeS, respectively).}
\end{figure}

To induce rotation by means of the torsion in the biphenyl-like molecule~[Fig.~\ref{fig:fig1}(a)], we have to induce a large torsional motion. Experimentally, the torsional motion can be initiated using non-resonant lasers. A long ns pulse is used to align the molecule, and then a short ps pulse kicks the planar groups, causing the torsional dynamics. The measurements, however, show that this procedure leads to oscillations in the dihedral angle [$2 \rho$ in Fig.~\ref{fig:fig1}(a)] only a few degrees from the equilibrium position~\cite{Madsen2009,Christensen2014}. Theoretical studies using a single laser lead to the same conclusion~\cite{Coudert2015}. We propose a different scenario, consisting of changing the torsional potential of the molecule by means of an electronic excitation. We select two designed torsional potentials to provoke different torsional dynamics, covering the behavior induced by a coplanar and a perpendicular relative orientation of the phenyl rings. Depending on the electronic state, the torsional wavepacket propagates in different regions, allowing us to explore several configurations of the molecule. The process is the following: i) the molecule is aligned with a ns pulse. We assume that the switching-on of the laser is adiabatic and that the torsional-rotational wavefunction is an eigenstate of the field-dressed Hamiltonian, ii) a short resonant laser excites the ground state potential electronic surface to an excited electronic state of the molecule. We approximate this process as instantaneous, because it is much faster than the torsional time scale of the molecule. Then we can use the sudden approximation and propagate the torsional wavepacket in this new torsional potential. 
 The scheme of this process is shown in Fig.~\ref{fig:fig1}(b). We consider three electronic states of the molecule~\cite{Ortigoso2013,Coudert2015}: the ground state potential energy surface (GS), the parallel state (PaS) and perpendicular state (PeS), which are described in detail in Sec.~\ref{sec:the_hamiltonian}. Briefly, in the PaS the internal potential has a minimum for the coplanar configuration, whereas the minimum in PeS is in the perpendicular configuration, a situation mimicking, for example, the nature of the states considered in Refs.~\cite{Baca1979,Akiyama1986,Im1988}. Clearly a transition from GS to either PaS or PeS will induce a large amplitude torsional motion. 

\subsection{The Hamiltonian}
\label{sec:the_hamiltonian}
In this section we describe the degrees of freedom of the system as well as the Hamiltonian. We consider the rotational and torsional dynamics of a biphenyl molecule as a prototype of a biphenyl-like molecule  formed by two phenyl rings, and which can rotate around a principal axis of inertia with a given internal-rotation barrier [Fig.~\ref{fig:fig1}(a)]. We work in the framework of the Born-Oppenheimer and the rigid rotor approximation except for the torsion between each ring of the molecule. % In particular, we describe the dynamics of the molecule interacting with a non-resonant laser field after an electronic transition induced by a pump laser.
% We consider a non-resonant laser field with a frequency much larger than the inverse of the typical time scale of the system.
In order to describe the orientation of a solid rigid molecule in space, we use the Euler angles $\Omega=(\phi,\,\theta,\,\chi)$, which relate the $XYZ$ axes of the laboratory fixed frame (LFF) with the $xyz$ axes of the MFF~\cite{Zare1988}. To describe the torsion of the molecule, we use the angle $\rho$, which shows an internal-rotation of $-\rho$ of one ring and $\rho$ of the other. Hence the dihedral angle between the two rings is $2\rho$. This definition of the torsion does not affect the principal axes of inertia, what simplifies the expression of the Hamiltonian~\cite{Merer1973}. The Hamiltonian of the system reads
\begin{equation}
\operator{H}=\operator{H}_\textup{tors-rot}+\operator{H}_\textup{int}(t),
  \label{eq:hamiltonian_total}
\end{equation}
where $\operator{H}_\textup{tors-rot}$ is the field-free Hamiltonian corresponding to the torsional-rotational dynamics and $H_\textup{int}(t)$ is the interaction with the non-resonant laser field. The torsional-rotational term reads~\cite{Merer1973}
\begin{eqnarray}
\nonumber 
\operator{H}_\textup{tors-rot}&=&A\operator{P}^2+A\operator{J}_z^2+\cfrac{1}{2}\left(B_x(\rho)+B_y(\rho)\right)\left(\operator{\mathbf{J}}^2-\operator{J}_z^2\right)\\
  \label{eq:hamiltonian_tors_rot}
&&+\cfrac{1}{4}\left(B_x(\rho)-B_y(\rho)\right)\left(\operator{J}_+^2+\operator{J}_-^2\right)+V(\rho),
\end{eqnarray}
where $\operator{P}=-i\frac{\partial}{\partial \rho}$ describes the torsional motion and $\operator{J}_k$ with $k=x,\,y,\,z$ are the angular momentum operators the rotational degrees of freedom and $\operator{J}_+$ and $\operator{J}_-$ are the corresponding ladder operators. Note that $\operator{\mathbf{J}}^2-\operator{J}_z^2=\operator{J}_x^2+\operator{J}_y^2$ and $\cfrac{1}{2}\left(\operator{J}_+^2+\operator{J}_-^2\right)=\operator{J}_x^2-\operator{J}_y^2$. In Eq.~\eqref{eq:hamiltonian_tors_rot}, we take $A=0.095833~\text{cm}^{-1}$ to describe the rotational constant around the $z$ axis of the MFF~\cite{Coudert2015} and $B_x(\rho)$ and $B_y(\rho)$ are the rotational constants around the $x$ and $y$ axes of the MFF. They can be expressed as  functions of $\rho$~\cite{Coudert2015} 
\begin{eqnarray}
\label{eq:bx_plus_by}
    \cfrac{1}{2}\left(B_x(\rho)+B_y(\rho)\right)&=&\cfrac{B}{1-\frac{B^2}{4A^2}\cos^22\rho},\\
\label{eq:bx_minus_by}                                      
    \cfrac{1}{2}\left(B_x(\rho)-B_y(\rho)\right)&=&B^2\cfrac{\cos2\rho}{2A(1-\frac{B^2}{4A^2}\cos^22\rho)},
\end{eqnarray}
with $B={0.016952}~\text{cm}^{-1}$.
Finally, $V(\rho)$ is the internal torsional potential, which may be written as~\cite{Coudert2015}
\begin{equation}
V(\rho)=V_0-\cfrac{1}{2}\sum_{n=1}^\nu V_{4n}\cos(4n\rho).
  \label{eq:v_torsion}
\end{equation}
In Table~\ref{table:parameters_ges_pas_pes} we show the parameters of the torsional potentials of GS, PaS and PeS~[see Fig~\ref{fig:fig1}].
\begin{table}[ht]
    \caption{\label{table:parameters_ges_pas_pes} Parameters $\left(\text{in units of cm}^{-1}\right)$ of the internal potentials of Eq.~\eqref{eq:v_torsion} for GS, PaS and PeS~\cite{Coudert2015}.}
  \begin{ruledtabular}
    \begin{tabular}{cccc}
& GS & PaS & PeS \\
\hline
&  &  &  \\
$V_0$ &345.340&250&250\\
$V_4$ &-32.288&500&-500\\
$V_8$ &-738.683&0&0\\
$V_{16}$ &-136.964&0&0\\
$V_{20}$ &-84.920&0&0\\
$V_{24}$ &-41.0610&0&0\\
$V_{28}$ & -32.562&0&0\\
    \end{tabular}
  \end{ruledtabular}
\end{table}
These potentials represent the main properties and shapes of the potential curves for biphenyl~\cite{Beenken2005}. We consider the interaction with a non-resonant linearly polarized laser along the $Z$ axis of the LFF, that does not affect the electronic or vibrational structure of the molecule. 
%Assuming that the angular frequency of the laser is much larger than the inverse of the rotational or torsional period, we can average over the rapid oscillation, to obtain
The effective interaction Hamiltonian reads~\cite{stapelfeldt:rev_mod_phys_75_543}
\begin{eqnarray}
  \operator{H}_\textup{int}&=& -\cfrac{1}{4}E_L(t)^\dagger\tensornote{\alpha}(\rho)E_L(t)=\\
&=& -\cfrac{E_L(t)^2}{8}\left[(2\alpha_z(\rho)-\alpha_y(\rho)-\alpha_x(\rho))\cos^2\theta+\right.
\nonumber\\
 \label{eq:h_int_linear_normal} 
&&\left.\alpha_x(\rho)+ \alpha_y(\rho)-(\alpha_y(\rho)-\alpha_x(\rho))\sin^2\theta\cos 2\chi\right]
\nonumber
\end{eqnarray}
where  $E_L(t)$ is the envelope of the electric field of the laser and $\tensornote{\alpha}(\rho)$ is the static polarizability tensor of the molecule. In the frame defined by the principal axes of inertia $\tensornote{\alpha}(\rho)$ is diagonal with matrix elements $\alpha_x(\rho)$,~$\alpha_y(\rho)$ and~$\alpha_z(\rho)$. By approximating the full polarizability,~$\tensornote{\alpha}(\rho)$, by the sum of the contributions from the two rings of the biphenyl, determined by the rotation operator $R(\rho)$ (see Appendix~\ref{sec:polarizability_molecule}), we obtain
\begin{equation}
  \label{eq:total_alpha}
  \tensornote{\alpha}(\rho)=R(\rho)\tensornote{\alpha}^0 R^{-1}(\rho)+R(-\rho)\tensornote{\alpha}^0 R^{-1}(-\rho).
\end{equation}
 Then, the elements of $\tensornote{\alpha}(\rho)$ can be written as
\begin{eqnarray}
  \label{eq:polarizability_x}
\alpha_x(\rho)&=&2(\alpha_x^0\cos^2\rho+\alpha_y^0\sin^2\rho)\\
  \label{eq:polarizability_y}
\alpha_y(\rho)&=&2(\alpha_x^0\sin^2\rho+\alpha_y^0\cos^2\rho)\\
  \label{eq:polarizability_z}
\alpha_z(\rho)&=&2\alpha_z^0.
\end{eqnarray}
Substituting Eqs.~\eqref{eq:polarizability_x}-\eqref{eq:polarizability_z} in Eq.~\eqref{eq:h_int_linear_normal}, the interaction with the laser can be written as~\cite{Coudert2011}
\begin{eqnarray}
 \nonumber
  \operator{H}_\textup{int}% &=&\cfrac{E_C^2}{8}\left[ (2\alpha_z^0-\alpha_y^0-\alpha_x^0)\cos^2\theta-2\alpha_z^0-\alpha_x^0-\alpha_y^0\right.\\
%  \label{eq:h_int_circular}
% &&\left.-(\alpha_y^0-\alpha_x^0)\sin^2\theta\cos 2\chi\cos 2\rho\right]+\\
% \nonumber
&=& -\cfrac{E_L(t)^2}{4}\left[(2\alpha_z^0-\alpha_y^0-\alpha_x^0)\cos^2\theta+\alpha_x^0+\alpha_y^0+\right.\\
 \label{eq:h_int_linear}
&&\left.-(\alpha_y^0-\alpha_x^0)\sin^2\theta\cos 2\chi\cos 2\rho\right],
\end{eqnarray}
where $\alpha_k^0$ with $k=x,\,y,\,z$ are the diagonal components of the polarizability tensor of each phenyl ring, being $\alpha_x^0=\alpha_z^0=11.7~\text{\AA}^3~\text{and}~\alpha_y^0=7.1~\text{\AA}^3$~\cite{Coudert2015}. 
\subsection{Symmetry considerations and basis set expansion}
\label{sec:symmetry_considerations}
The full description of the eigenstates and the wavefunctions requires consideration of the symmetry properties of the system and its associated representations. Briefly, the symmetry group corresponding to the biphenyl-like molecule in Fig.~\ref{fig:fig1}(a) is $G_{16}^2$ of the Longuet-Higgins permutation group~\cite{Longuet-Higgins1963}, formed by the identity $E$, the two-fold rotations around the $x$ and $y$ axis of the MFF, $C_2(x)$ and $C_2(y)$, four-fold rotations around the $z$ axis, $C_4(z)$, and its square, $\left[C_4(z)\right]^2$. The effect on the Euler angles and the torsional angle of these generating operations are collected in Table~\ref{table:generating_operations_g16_2}.
\begin{table}[ht]
    \caption{\label{table:generating_operations_g16_2} Generating operations of the symmetry group $G_{16}^2$(see Ref.~\cite{Merer1973}), which describes the torsional-rotational symmetry operations of the biphenyl-like molecule of Fig.~\ref{fig:fig1}(a).}    
  \begin{ruledtabular}
    \begin{tabular}{cccccc}
      & $E$& $C_4(z)$ & $C_2(x)$ & $C_2(y)$ & $\left[C_4(z)\right]^2$  \\
\hline
&&&&&\\
      $\phi\rightarrow$ & $\phi$& $\phi$ & $\phi+\pi$ & $\phi+\pi$ & $\phi$\\
      &&  &  &  &  \\
      $\theta\rightarrow$ & $\theta$& $\theta$ & $\pi-\theta$ & $\pi-\theta$ & $\theta$\\
      &&  &  &  &  \\
      $\chi\rightarrow$ & $\chi$ & $\chi+\frac{\pi}{2}$ & $-\chi$ & $-\chi$ & $\chi+\pi$\\
      &&  &  &  &  \\
      $\rho\rightarrow$ &$\rho$& $\frac{\pi}{2}-\rho$ & $\rho$ & $\pi-\rho$ & $\rho+\pi$\\
    \end{tabular}
  \end{ruledtabular}
\end{table}
The representations $\Gamma$ corresponding to this symmetry group are $A_{1g}^+,\,A_{2u}^+,\,A_{1u}^-,\,A_{2g}^-,\,B_{1g}^+,\,B_{2u}^+,\,B_{1u}^-,\,B_{2g}^-,\,E^+$ and $E^-$, where the first 8 representations are one-dimensional, and the latter are two-dimensional representation. In this notation, $A$ and $B$ refer to even and odd parity under $C_4(z)$; $1$ and $2$ refer to even and odd parity under $C_2(x)$; $+$ and $-$ to even and odd parity under  $C_2(y)$ and $g$ and $u$ to even and odd parity under $C_2(x)C_2(y)$~\cite{Merer1973,Longuet-Higgins1963}. In this study we describe states belonging to the torsional-rotational representation  $\Gamma_\textup{t-r}=A_{1g}^+$ which contains the torsional-rotational ground state. $\Gamma_\textup{t-r}$ expands as a sum of direct products of torsional ($\Gamma_\textup{tors}$) and rotational representations ($\Gamma_\textup{rot}$). The representation $\Gamma_\textup{t-r}=A_{1g}^+$ can be decomposed as $A_{1g}^+\times A_{1g}^++B_{1g}^+\times B_{1g}^+$~\cite{Coudert2011}. Then, to construct the basis set functions of $\Gamma_\textup{t-r}=A_{1g}^+$ we use the basis set functions of $\Gamma_\textup{rot}$ and $\Gamma_\textup{tors}$. For the rotational part, we use a basis set formed by the Wang functions, $|JKM\pm\rangle_w$, defined as~\cite{Hougen1962}
\begin{eqnarray}
\label{eq:wang_states_k_ne_0}
    |JKM\pm\rangle_w&=&\cfrac{1}{\sqrt{2}}\left(|JKM\rangle\pm |J-KM\rangle\right),\, K>0,    \\
\label{eq:wang_states_k_eq_0}
    |J0M+\rangle_w&=&|J0M\rangle,\, K=0.
\end{eqnarray}
labeled by the total angular momentum number, $J$, its projection along the $z$-axis of the MFF, $K$, its projection along the $Z$-axis of the LFF, $M$, and the parity, $(\pm)$. In Eqs.~\eqref{eq:wang_states_k_ne_0} and~\eqref{eq:wang_states_k_eq_0} $\ket{JKM}$ are the field-free symmetric rigid rotor eigenstates~\cite{Zare1988}
\begin{equation}
\langle \Omega | J K M \rangle =(-1)^{M-K}\sqrt{\cfrac{2J+1}{8\pi^2}}D_{-M,-K}^J(\Omega) %=\sqrt{\cfrac{2J+1}{8\pi^2}}D_{M,K}^{J}(\Omega)^*,
  \label{eq:ff_eigenfunctions}
\end{equation}
with $D_{M,K}^{J}(\Omega)$ the Wigner matrix elements. The torsional basis functions can be expanded in a discrete variable representation (DVR), $u_\alpha(\rho)$, derived from properly symmetrized cosine and sine functions in the torsional angle~[see also Sec.~\ref{sec:numerical_methods} and Appendix~\ref{sec:matrix_elements}]. In Table~\ref{table:basis_representation} we summarize the labels of the Wang functions and the torsional functions needed to build the DVR.
\begin{table}[ht]
    \caption{\label{table:basis_representation} Basis set functions of $\Gamma_\textup{rot}$ and $\Gamma_\textup{tors}$ needed to construct the basis set of $\Gamma_\textup{t-r}=A_{1g}^+$. $J$ and $K$ denote the total angular momentum and its projection on the $z$ axis of the MFF, respectively, and $K_t$ is the angular momentum corresponding to the torsion,~\ie, corresponding to the operator~$\operator{P}=-i\frac{\partial}{\partial \rho}$ in Eq.~\eqref{eq:hamiltonian_tors_rot}.}    
  \begin{ruledtabular}
    \begin{tabular}{ccc}
      $\Gamma_\textup{t-r}$  &$\Gamma_\textup{rot}$  & $\Gamma_\textup{tors}$ \\
                             &&\\
                             & $\left(\mod(J,2),\mod(K,4),\pm\right)$   & Functions  \\
      \hline
      $A_{1g}^+$ & $(0,0,+),(1,0,-)$   & $\cfrac{1}{\sqrt{2\pi}},\quad K_t=0$ \\
                             & & $\cfrac{1}{\sqrt{\pi}}\cos K_t \rho,\quad K_t=4,8\ldots $ \\
      \hline
      $B_{1g}^+$ & $(0,2,+),(1,2,-)$   & $\cfrac{1}{\sqrt{\pi}}\cos K_t \rho,\quad K_t=2,6\ldots $ \\
    \end{tabular}
  \end{ruledtabular}
\end{table}
% Its construction and associated matrix elements are described in detail in Appendix~\ref{sec:symmetries_and_representations_basis_sets}.

\section{Numerical methods}
\label{sec:numerical_methods}

The wavefunction is expanded as a linear combination of products of symmetrized rotational field-free eigenstates times DVR functions in the torsional coordinate
\begin{eqnarray}
\nonumber
&&\psi_M(\Omega,\rho,t)=\\
\label{eq:expansion_coef}
&&\sum_{J=0}^{J_\textup{max}}\sum_{K=-J}^J \sum_{\alpha=1}^{n_\textup{max}}c_{JKMn}(t) \langle \Omega|JKM\pm\rangle_w u_\alpha(\rho),
\end{eqnarray}
where we consider the restrictions in the sum and in the parity of the Wang states given by the representation $\Gamma_\textup{rot}$ collected in Table~\ref{table:basis_representation}. Note that $M$ is a good qunatum number because the Hamiltonian commutes with arbitrary rotations around the polarization axis of the laser, i.e., the $Z$ axis of the LFF. The DVR functions are not chosen symmetrized, because it would imply working with two different DVR grids. In our case, the basis set in the representation $A_{1g}^+$ is built using the functions $\cos K_t\rho$ with $\mod(K_t,4)=0$ ($=A_{1g}^+$) and 2 ($B_{1g}^+$), see Table~\ref{table:basis_representation}, which leads to two different DVRs. By $K_t$ we denote the angular momentum of the torsion. Then, our strategy consists of using the minimum DVR common to both basis sets, $\cos K_t\rho$ with even $K_t$. Finally, to identify the symmetry of a given function we use the projector on $\Gamma_\textup{t-r}=A_{1g}^+$, $\mathbb{P}=\sum \ket{\psi_j}\bra{\psi_j}$, with $\ket{\psi_j}\in A_{1g}^+$.

The process is simulated as follows: i) We obtain a field-dressed state in the presence of the non-resonant linearly polarized laser field by diagonalizing the field-dressed Hamiltonian using the ARPACK library~\cite{arpack,Lehoucq1997}. ii) We take the projection of the eigenstates on $\Gamma_\textup{t-r}=A_{1g}^+$, and select the initial state. iii) Finally, we change the torsional potential $V(\rho)$ to the one belonging to the excited electronic state, and the wavepacket is propagated in this new Hamiltonian using the short iterative Lanczos method~\cite{mctdh}. To reach converged results we need to use $J_\textup{max}=30$ and $n_\textup{max}=51$.

\section{Results}
\label{sec:results}

In this section we analyze the implications of the electronic transition from GS to PaS or PeS~[Fig.~\ref{fig:fig1}] on the torsional and rotational dynamics for a given field-dressed state at the peak of a linearly polarized Gaussian laser. We use Gaussian pulses $I(t)=I_0\exp\left(-4\ln 2 t^2/\tau^2\right)$ for several intensities and a fixed FWHM of $\tau=10$~ns. The degree of alignment is $\expected{\cos^2\theta}$, which ranges from 0 if the $z$ axis of the MFF is contained in the $XY$ plane of the LFF to 1 for total alignment along the $Z$ axis of the LFF,~\ie, $\theta$ is the angle between $Z$ and $z$. The total angular momentum and its projection on the $z$ axis of the MFF, $\expected{J^2}$ and $\expected{K^2}$, are used to evaluate the impact of the torsional dynamics on the overall rotation of the molecule. Finally,  to describe the evolution of the torsional angle, $\rho$, we define the quantity $\rho_e=\arccos\left(\expected{\cos 4\rho}\right)/4$.%Finally,  $\rho_e=\arccos\left(\expected{\cos 4\rho}\right)/4$, is a torsional average value of the half of the angle formed by the two planar groups.

\subsection{Ground state to planar configuration}
\label{sec:ground_state_to_planar_configuration}

\begin{figure}
  \centering
  \includegraphics[width=\linewidth]{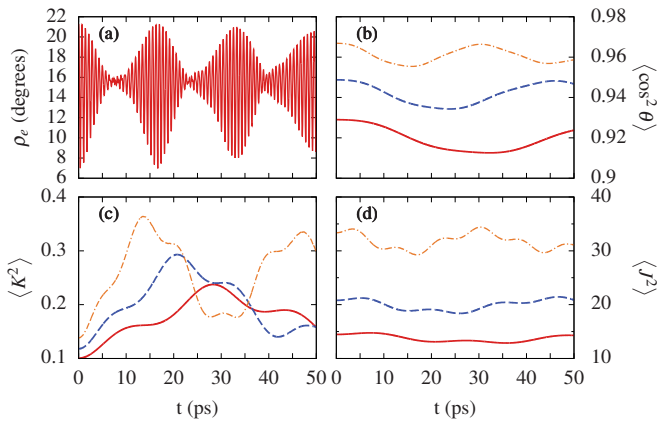}
  \caption{\label{fig:fig2} (a) $\rho_e$, (b) $\expected{\cos\power{2}\theta}$, (c) $\expected{K\power{2}}$ and (d) $\expected{J\power{2}}$ for the transition from GS to PaS starting in the ground state of $A_{1g}^+$ for $I_0=10^{11}~\text{(solid red)}~,2\times 10^{11}~\text{(dashed blue) and}~5\times 10^{11}$~W/cm${}^{-2}$ (dotted orange). The quantities in (c) and (d) are in units of $\hbar^2$.}  
\end{figure}

We first analyze the dynamics of the torsional-rotational low-lying eigenstates of the biphenyl-like molecule after the electronic transition from GS to PaS. After this process, the wavepacket moves to the minimum of the potential at $\rho=0~\text{and}~90\degree$ driven by the slope of the new torsional potential, see Fig.~\ref{fig:fig1}(b).  This dynamics leads to a change in the rotational parameters~$B_x(\rho)$ and~$B_y(\rho)$ which provokes in turn a change in the overall rotational dynamics. 
In Figure~\ref{fig:fig2} we show the resulting effect for the torsional-rotational ground state in GS as the initial state for several intensities of the aligning lasers.
The torsional dynamics is indistinguishable for $I_0=1,~2~\text{and}~5\times 10^{11}~\text{W}/\text{cm}^{-2}$, as we see in Fig.~\ref{fig:fig2}(a). The average torsional angle is $\rho_e\approx 21.4\degree$ at $t=0$ and rapidly decreases to $\approx 7\degree$ at $t=0.35~\text{ps}$. Then the amplitude of the oscillation is decreasing until $t\approx 7~\text{ps}$ where $\rho_e\approx 15.6\degree$ is reached, which is also the mean value during the rest of the propagation. Then, the amplitude increases again, and a similar lobe structure is formed. As we continue the propagation, the minimum amplitude is larger at the beginning of the lobe due to interference of the torsional wavepackets. For example, the fourth lobe starts at $t\approx 40~\text{ps}$; the average value remains constant, but the amplitude is larger than for the previous lobes.
The torsional dynamics has, however, a strong impact on the overall rotation of the molecule, as revealed in the expectation values. The alignment, $\expected{\cos^2\theta}$, of the $z$ axis of the MFF, which is the most polarizable axis of the molecule, the projection of the angular momentum along the molecular $z$ axis $~\expected{K^2}$ and the total angular momentum $\expected{J^2}$, are shown in Figs.~\ref{fig:fig2}(b)-(d).  The alignment $\expected{\cos^2\theta}$  shows a weak irregular oscillatory behavior with a frequency which depends on the peak intensity $I_0$. This relationship depends on the energy gap among the field-dressed states, which changes with the laser interaction and would be different for a molecule with other rotational constants. For all the peak intensities considered here, the alignment decreases just after the electronic transition has occurred. Moreover, the difference between maximum and minimum decreases with increasing intensity of the laser, being $\approx 0.016$ for $I_0=10^{11}$~W/cm${}^{-2}$ and $\approx 0.0115$ for $I_0=5\times 10^{11}$~W/cm${}^{-2}$, because the most polarizable axis of the molecule is more tight to the laser polarization axis for the more intense laser.

Figure ~\ref{fig:fig2}(c) shows that $\expected{K^2}$ increases from $t=0$, reaches a peak and continues with irregular oscillations with an amplitude increasing with $I_0$. These oscillations are influenced by more than one frequency.  This can be understood in terms of the rotational constants associated to PaS; for the minimum in the PaS, that is, $\rho=0,90\degree$, the asymmetry of the molecule $\frac{1}{2} 
|\left(B_x(\rho)-B_y(\rho)\right) |= 2AB^2/(4A^2-B^2)$ is maximum, see Eq.~\eqref{eq:bx_minus_by}. In the Hamiltonian of Eq.~\eqref{eq:hamiltonian_tors_rot} this term is responsible of the mixing of $K$s, \ie, $K$ is not a good quantum number. Then, after the induced transition, the population increases in states with different contributions of $K$, leading to couplings among different states. By comparing Figs.~\ref{fig:fig2}(b) and (c), we see that the main frequencies in $\expected{\cos^2\theta}$ and $\expected{K^2}$ are similar. Furthermore, note that a minimum in $\expected{K^2}$ corresponds to a maximum in $\expected{\cos^2\theta}$ and vice versa. Since the alignment is weakly affected on the natural time scale of the torsion, we can conclude that $\expected{K^2}$ is driven by the overall rotation. 

Finally, we see that the oscillations in $\expected{J^2}$ are a combination of those in  $\expected{K^2}$ and $\expected{\cos^2\theta}$, since the main trend corresponds to the alignment but it is modulated by the internal rotation of the molecule. In fact, $J$ acts as a bridge between $K$ and the alignment, because $K$ is an important contribution to the total angular momentum $J$, which drives the alignment of the $z$ axis of the MFF~\cite{Zare1988}. 
\begin{figure}
  \centering
  \includegraphics[width=\linewidth]{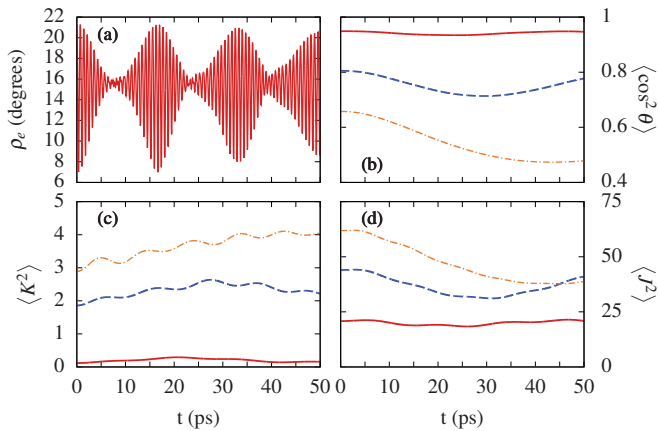}
  \caption{\label{fig:fig3} (a) $\rho_e$, (b) $\expected{\cos\power{2}\theta}$, (c) $\expected{K\power{2}}$ and (d) $\expected{J\power{2}}$ for the transition from GS to PaS starting in the ground state (solid red), first (dashed blue) and second (dotted orange) excited states of $A_{1g}^+$ for $I_0=2\times 10^{11}$~W/cm${}^{-2}$. The quantities in (c) and (d) are in units of $\hbar^2$.}  
\end{figure}

Now we turn to an analysis of the torsional-rotational dynamics for the three lowest-lying energy levels of the GS for $I_0=2\times 10^{11}$~W/cm${}^{-2}$. In Fig.~\ref{fig:fig3}, we see that the torsional dynamics are indistinguishable for the three states, [Fig.~\ref{fig:fig3}(a)]. The alignment also decreases from $t=0$ and starts to librate for the states considered in Fig.~\ref{fig:fig3}(b). For $t=0$ the alignment is smaller for excited states,  therefore, the overall rotation and the alignment are more sensitive to the induced torsional dynamics. In particular, the initial alignment of the second state is $\approx 0.805$ and it decreases to a minimum of $0.713$ at $\approx 29.625$~ps. For the third state the behavior follows the same decreasing pattern at the beginning. Starting at $\approx 0.657$ it reaches $\approx 0.474$ at the minimum at $t\approx 43.5~\text{ps}$. In Fig.~\ref{fig:fig3}(d), we clearly see the signature of $\expected{K^2}$ and $\expected{\cos^2\theta}$ on $\expected{J^2}$, that is, the long oscillation comes from the alignment and the ripples are determined by the oscillations in $\expected{K^2}$.% In Fig.~\ref{fig:fig3}(c) we see that the oscillations are driven mainly by one frequency, and it is modulated by a weaker contribution which provoke the ripples in the pattern.
\subsection{Ground state to perpendicular configuration}
\label{sec:ground_state_to_perpendicular_configuration}

In this section we analyze the transition from the GS to the PeS, displayed in Fig.~\ref{fig:fig1}(b). Before the transition, the torsional wavepacket is divided in two distributions located at the minimum of the GS, whose overlap is negligible. After the electronic excitation the distributions are placed on the slope of the potential which drives them to the absolute minimum at $\rho=45\degree$.

\begin{figure}
  \centering
  \includegraphics[width=\linewidth]{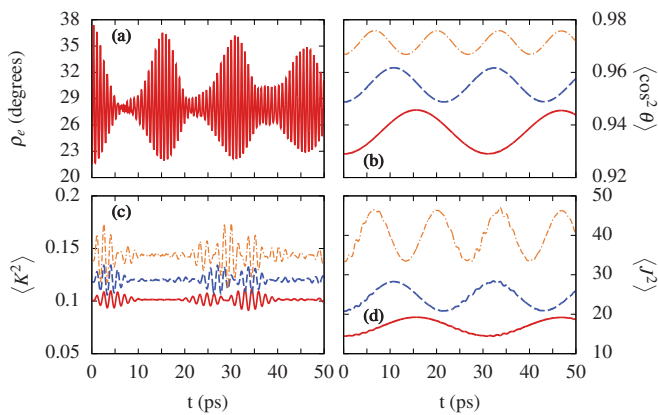}
  \caption{\label{fig:fig4} (a) $\rho_e$, (b) $\expected{\cos\power{2}\theta}$, (c) $\expected{K\power{2}}$ and (d) $\expected{J\power{2}}$ for the transition from GS to PeS starting in the ground state of $A_{1g}^+$ for $I_0=10^{11}~\text{(solid red)}~,2\times 10^{11}~\text{(dashed blue) and}~5\times 10^{11}$~W/cm${}^{-2}$ (dotted orange). The quantities in (c) and (d) are in units of $\hbar^2$.}  
\end{figure}

In Fig.~\ref{fig:fig4}, we show the dynamics for the torsional-rotational ground state for three intensities of the alignment pulse.
In Fig.~\ref{fig:fig4}(a) we see that the torsional angle $\rho_e\approx 21.4\degree$ starts at $t=0$ and increases until $\approx 37.4\degree$. This libration continues with smaller amplitudes until it reaches an average value of $\approx 28\degree$. Then, it starts increasing again, and another lobe is obtained. The amplitude in the successive lobes becomes more and more similar. As in the GS to PaS process, the torsional dynamics are indistinguishable for these three intensities. In contrast to the previous case, the alignment $\expected{\cos^2\theta}$ shows a regular oscillation trend and furthermore, the initial value is a lower bound during the propagation. We observe that the higher peak intensity of the laser, the larger  the frequency of the oscillation, and also its amplitude. We see in Fig.~\ref{fig:fig4}(d) that the oscillation of $\expected{J^2}$ coincides to the alignment (appropriated scaled), despite the small ripples in $\expected{J^2}$, which correspond to the strongest oscillations in $\expected{K^2}$. As we see in Fig.~\ref{fig:fig4}(c), the mean value of $\expected{K^2}$ is unaltered, since in the perfect perpendicular configuration, $\rho=45\degree$, the rotational Hamiltonian corresponds to a symmetric rotor, see Eq.~\eqref{eq:bx_minus_by}, \ie, it couples to other states $\Delta K=0$. In contrast to the transition from GS to PaS, the fluctuations in $K$ are not driven by the fluctuations in $\expected{\cos^2\theta}$ and $\expected{J^2}$, but they are located at the same interval as the first and third lobes of $\rho_e$. This means that, in contrast to the transition from GS to PaS, the coupling of different $K$'s is due mainly to the torsional dynamics. 
\begin{figure}
  \centering
  \includegraphics[width=.8\linewidth]{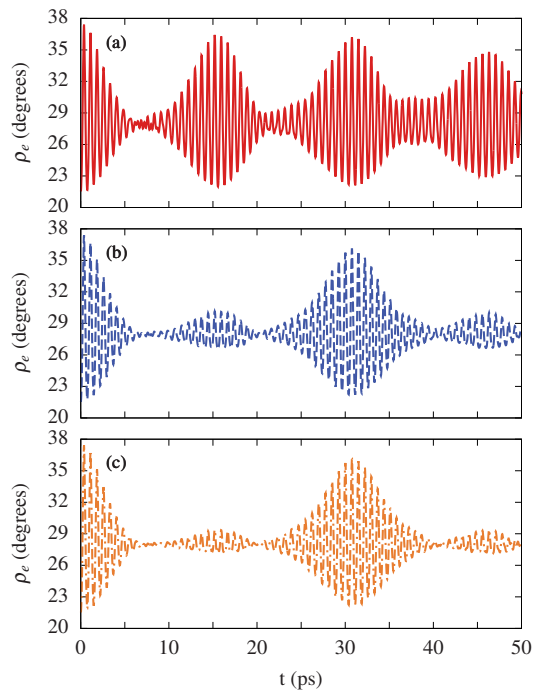}
  \caption{\label{fig:fig5} $\rho_e$ for (a) the ground state, (b) the first and (c) second excited states of $A_{1g}^+$ for the transition from GS to PeS for $I_0=2\times 10^{11}$~W/cm${}^{-2}$.}  
\end{figure}

\begin{figure}
  \centering
  \includegraphics[width=.8\linewidth]{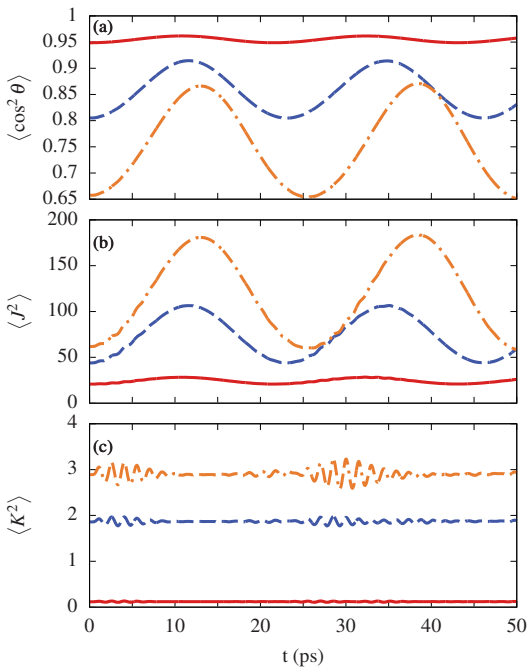}
  \caption{\label{fig:fig6} (a) $\left\langle\cos^2\theta\right\rangle$, (b) $\left\langle J^2\right\rangle$ and (c) $\left\langle K^2\right\rangle$ for the transition from GS to PeS for the ground state (solid red), the first (dashed blue) and (c) second (dotted orange) excited states of $A_{1g}^+$ for $I_0=2\times 10^{11}$ W/cm${}^{-2}$. The quantities in (b) and (c) are in units of $\hbar^2$.}  
\end{figure}

In Figs.~\ref{fig:fig5} and~\ref{fig:fig6} we show the expectation values corresponding to the propagation of the three lowest-lying states of the GS for $I_0=2\times 10^{11}$~W/cm${}^{-2}$. 
We observe in Fig.~\ref{fig:fig5} that the torsional dynamics differs for different states opposed to the GS to PaS transition.  In the present case, the first and third lobes are the same for the three states. The second and fourth lobes, however, have approximately the same amplitude as the other lobes for the ground state, but are strongly suppressed for the other two torsional rotational states. This suppression corresponds to the destructive interference of the torsional wavepackets derived by the energy level structure and the common minimum reached at $\rho=45\degree$ by the two parts of the wavepacket. 
In Fig.~\ref{fig:fig6} we show the rotational expectation values for these three states. At $t=0$, the alignment starts increasing for all the states shown in Fig.~\ref{fig:fig6}(a), and reaches a maximum value for $t\approx 10.8, 11.6~\text{and}~13$~ps. The amplitudes of the fluctuations depends on the state. For higher excited states, the amplitude is larger, since the kinetic rotational energy of the state can counteract the interaction of the laser. On the one hand, the ground state is weakly affected by the transition, since the alignment only increases $\approx 0.013$. On the other hand, there is a strong impact on the other states. For instance, at $t=0$, the alignment of the second state is $\approx 0.805$ and it increases up to $\approx 0.914$. Moreover, the improvement of the alignment is even larger for the third state, whose maximum alignment reached by the third state is $\approx 0.866$, with $\approx 0.657$ its value at $t=0$.

The total angular momentum $\expected{J^2}$ in Fig.~\ref{fig:fig6}(b) follows the same oscillatory pattern as the alignment. The mixing of the $J$ is very strong for the second and the third state, going from $\approx 61.86$ to $\approx 180.87$ for the last one. There are small ripples in the $\expected{J^2}$ that can not be observed in the alignment. These ripples correspond to the fluctuations in the expectation value of $\expected{K^2}$ [Fig.~\ref{fig:fig6}(c)] which are located in the intervals $t=0$ to $10$~ps and from $20$ to $40$~ps for the three states. The reason is that $\expected{K^2}$ is only affected by the torsional dynamics, in contrast to the GS to PaS transition which is the same for the three states. %However, the amplitude of $\expected{K^2}$ is not the same for these states, being larger for the excited states.

\section{Conclusions and outlook}
\label{sec:conclusions_and_outlook}

In this paper we have analyzed the coupling of torsional and rotational dynamics of a non-rigid symmetric molecule, taken as a biphenyl-like molecule as an illustrative case. We showed that it is feasible to probe the rotational dynamics by inducing large amplitude torsional motion in an aligned molecule. To do so, we provoked a transition between two electronic levels, which are characterized by different internal torsional potentials~\cite{Baca1979}, inducing the propagation of the torsional wavepacket. We presented two different scenarios: the transition from the torsion-rotational ground state to a planar and to a perpendicular excited state configuration~\cite{Coudert2015}. The latter two were modelled to capture generic effects on the dynamics such that the conclusions of this work can be extrapolated to similar potentials of non-rigid molecules~\cite{Imamura1968,Beenken2005}.

We found that the torsional dynamics is not affected by the strength of the alignment laser, but depends on the torsion-rotational state. We showed that the torsional motion alters the rotational parameters, producing the change in the rotational motion. Depending on the final electronic state, the alignment can increase or decrease, being more pronounced for weaker laser fields and for the more excited states of the GS analyzed here. The angular momentum $J$, which characterizes the overall rotation, presents the same pattern as the alignment, whereas the internal rotation $K$ is modulated by either the overall rotation or the internal torsion, or by both, depending on the final electronic state. It should be possible to address experimentally the present theoretical findings on the torsional and rotational dynamics of non-rigid molecules for the laser parameters discussed in this paper. The techniques to measure the alignment and the torsional angle are available~\cite{Madsen2009a,Madsen2009,Hansen2012,Christensen2014}.

\appendix

\section{Calculation of the polarizability of the molecule}
\label{sec:polarizability_molecule}
In this appendix we derive the expression of the polarizability tensor $\tensornote{\alpha}(\rho)$ used in Sec.~\ref{sec:the_hamiltonian}. We assume that the polarizability tensor $\tensornote{\alpha}(\rho)$ of the biphenyl-like molecule is the sum of the polarizability tensor of the two rings, $\tensornote{\alpha}^0$. For $\rho=0$, $\tensornote{\alpha}^0$ is diagonal
\begin{equation*}
  \label{eq:alpha0_rho_0}
  \tensornote{\alpha}^0=\left(
  \begin{array}{ccc}
    \alpha_x^0 & 0 & 0\\
    0 & \alpha_y^0 & 0\\
    0 & 0 & \alpha_z^0 \\
  \end{array}\right).
\end{equation*}
First, we calculate the polarizability for each ring independently. The first ring is rotated an angle $\rho$ around the $z$ axis of the MFF
\begin{eqnarray}
\nonumber
&&R(\rho)\tensornote{\alpha}^0 R^{-1}(\rho)=\\
\nonumber
&&\left(
\begin{array}{ccc}
\alpha_x^0\cos^2\rho+\alpha_y^0\sin^2\rho  &(\alpha_y^0-\alpha_x^0)\sin\rho\cos\rho&0\\
 (\alpha_y^0-\alpha_x^0)\sin\rho\cos\rho&\alpha_y^0\cos^2\rho+\alpha_x^0\sin^2\rho&0\\
0 & 0& \alpha_z^0
\end{array}
\right),
\end{eqnarray}
and the other ring an angle $-\rho$
\begin{eqnarray}
\nonumber
&&R(-\rho)\tensornote{\alpha}^0 R^{-1}(-\rho)=\\
\nonumber
&&\left(
\begin{array}{ccc}
\alpha_x^0\cos^2\rho+\alpha_y^0\sin^2\rho  &(-\alpha_y^0+\alpha_x^0)\sin\rho\cos\rho&0\\
 (-\alpha_y^0+\alpha_x^0)\sin\rho\cos\rho&\alpha_y^0\cos^2\rho+\alpha_x^0\sin^2\rho&0\\
0 & 0& \alpha_z^0
\end{array}
\right),
\end{eqnarray}
where 
\begin{equation*}
  R(\rho)=\left(
    \begin{array}{ccc}
      \cos\rho&\sin\rho&0\\
      -\sin\rho&\cos\rho&0\\
      0 & 0& 1
    \end{array}
\right).
\end{equation*}
Finally, we obtain at the present level of approximation the total polarizability as the sum of the two terms
\begin{eqnarray*}
&&\tensornote{\alpha}(\rho)=R(\rho)\tensornote{\alpha}^0 R^{-1}(\rho)+R(-\rho)\tensornote{\alpha}^0 R^{-1}(-\rho)=\\
&&2\left(\begin{array}{ccc}
\alpha_x^0\cos^2\rho+\alpha_y^0\sin^2\rho  & 0 &0\\
0 &\alpha_y^0\cos^2\rho+\alpha_x^0\sin^2\rho&0\\
0 & 0& \alpha_z^0
\end{array}\right)
\end{eqnarray*}

%Note that we use $\rho$ instead of $2\rho$ or $4\rho$ for convention reasons~\cite{Coudert2015}.

\section{Hamiltonian matrix elements and basis set}
\label{sec:matrix_elements}
In this appendix, we discuss general aspects of the Hamiltonian matrix elements and the properties of the basis set used in Secs.~\ref{sec:the_hamiltonian} and~\ref{sec:numerical_methods}. The term in the Hamiltonian describing the interaction with the linearly polarized laser field in Eq.~\eqref{eq:h_int_linear} can be rewritten as a linear combination of the Wigner matrix elements, $D_{M,K}^{J}(\Omega)$~\cite{Zare1988}, 
  \begin{eqnarray}
\nonumber
    H_{\textup{int},L}(\rho,\Omega)&=&-\cfrac{E_L^2}{2}\left[\cfrac{\alpha_x\power{0}+\alpha_y\power{0}+\alpha_z\power{0}}{3}\right.\\
\nonumber
&&+\cfrac{2\alpha_z\power{0}-\alpha_y\power{0}-\alpha_x\power{0}}{3}D_{00}\power{2}(\Omega)\\
\nonumber%\label{eq:hamiltonian_interaction_linear}
&&\left.-\cfrac{\alpha_y\power{0}-\alpha_x\power{0}}{\sqrt{6}}\left(D_{02}\power{2}(\Omega)+D_{0-2}\power{2}(\Omega)\right)\cos 2\rho\right]
  \end{eqnarray}
To evaluate the Hamiltonian matrix elements in the field-free symmetric rotor eigenfunctions we use that~\cite{Zare1988}
\begin{eqnarray}
\nonumber
&&\left\langle J K M\left|D_{pq}^r(\Omega)\right|J' K' M'\right\rangle  =\sqrt{(2J+1)(2J'+1)}\times\\
\nonumber
&&(-1)^{M'-K'}
 \begin{pmatrix}
   J & r & J'\\
K & q & -K'
 \end{pmatrix}
 \begin{pmatrix}
   J & r & J'\\
M & p & -M'
 \end{pmatrix},
\end{eqnarray}
where the parentheses denote $3\mathcal{J}$ symbols~\cite{Zare1988}.

The DVR functions $u_\alpha(\rho)$ in Eq.\eqref{eq:expansion_coef} fulfill the following conditions~\cite{Muckerman1990}
\begin{eqnarray*}
&&  u_\alpha(\rho)=\sqrt{\omega_\alpha}\sum_{n=1}^N\phi_n^*(\rho_\alpha)\phi_n(\rho),\\
&&  u_\alpha(\rho_\beta)=\cfrac{\delta_{\alpha,\beta}}{\sqrt{\omega_\alpha}},\\
&&\int u_\alpha*(\rho)u_\beta(\rho)V(\rho)\mathrm{d}\rho=V(\rho_\alpha)\delta_{\alpha,\beta},
\end{eqnarray*}
where $\omega_\alpha$ and $\rho_\alpha$ are the weights and nodes of the DVR and $\{\phi_n(\rho)\}$ is an orthonormal basis set. To impose the symmetrization in the DVR functions we set $\phi_n(\rho)$ to be a set of $N+1$ functions in the $A_{1g}^+$ and $B_{1g}^+$ representations collected in Table~\ref{table:basis_representation}, that is, $\phi_0(\rho)=\cfrac{1}{\sqrt{2\pi}}$ and $\phi_n(\rho)=\cfrac{1}{\sqrt{\pi}}\cos 2n\rho$, with $n=1,\ldots,N$. The explicit expressions for the weights and nodes of the quadrature corresponding to the cosine functions, denoted by superscript $c$, are 
\begin{equation*}
    \omega^c_\alpha=\cfrac{2\pi}{N+1}, \rho^c_\alpha=\cfrac{\pi(\alpha-1/2)}{2(N+1)}
\label{eq:weights_nodes_cos2}
\end{equation*}
with $1\le \alpha\le N+1$ and the DVR functions  are
\begin{eqnarray*}
u^c_\alpha(\rho)&=&\cfrac{1}{2\sqrt{2\pi(N+1)}}\left(\cfrac{\sin (2N+1)(\rho-\rho_\alpha)}{\sin (\rho-\rho_\alpha)}\right.\\
&&\left.+\cfrac{\sin (2N+1)(\rho+\rho_\alpha)}{\sin (\rho+\rho_\alpha)}\right)%\left(\cfrac{1}{2\pi}+\cfrac{1}{\pi}\sum_{n=1}^N\cos(2n\rho_\alpha)\cos(2n\rho)\right)%  \label{eq:dvr_cos}
\end{eqnarray*}
and the matrix elements of the torsional momentum square in this basis, $\operator{P}^2_{\alpha,\beta}=\melement{u^c_\alpha}{\operator{P}^2}{u^c_\beta}$
\begin{eqnarray*}
\nonumber
&&\operator{P}^2_{\alpha,\beta}=-\int_0^{2\pi} u^c_\alpha(\rho)\cfrac{\partial^2}{\partial \rho^2}u^c_\beta(\rho)\mathrm{d}\rho= \\
\nonumber
&& =2\left\{
 \begin{array}{l}
   (-1)^{\alpha-\beta}\left[\cfrac{1}{\sin^2\left(\frac{\pi(\alpha-\beta)}{2(N+1)}\right)}-\cfrac{1}{\sin^2\left(\frac{\pi(\alpha+\beta+1)}{2(N+1)}\right)}\right],\quad \beta\ne \alpha,\\
   \left[\cfrac{2N^2+4N+3}{3}-\cfrac{1}{\sin^2\left(\frac{\pi(2\alpha+1)}{2(N+1)}\right)}\right], \quad \beta=\alpha.
 \end{array}
\right.
\end{eqnarray*}

\begin{acknowledgments}
This work was supported by the ERC- StG (Project No. 277767-TDMET) and the Villum Kann Rasmussen (VKR) Center of Excellence QUSCOPE. The numerical results presented in this work were obtained at the Centre for Scientific Computing, Aarhus and Proteus Scientific Computing Cloud, Universidad de Granada (Spain).
\end{acknowledgments}
%\bibliographystyle{apsrev4-1}
%\bibliography{torsion_molecules_superrotors}

%merlin.mbs apsrev4-1.bst 2010-07-25 4.21a (PWD, AO, DPC) hacked
%Control: key (0)
%Control: author (0) dotless jnrlst
%Control: editor formatted (1) identically to author
%Control: production of article title (0) allowed
%Control: page (1) range
%Control: year (0) verbatim
%Control: production of eprint (0) enabled
%

\end{document}